\documentclass[12pt,a4paper]{article}
\usepackage{amsmath}
\usepackage{amssymb}
\usepackage{amsfonts}
\usepackage{graphicx}
\usepackage[onehalfspacing]{setspace}

\def\BE{\begin{equation}}
\def\EE{\end{equation}}
\def\BEA{\begin{eqnarray}}
\def\EEA{\end{eqnarray}}
\newtheorem{theorem}{Theorem}

\newtheorem{corollary}{Corollary}

\newtheorem{definition}{Definition}

\newtheorem{lemma}{Lemma}

\newtheorem{proposition}{Proposition}

\newenvironment{proof}[1][Proof]{\noindent\textbf{#1.} }{\ \rule{0.5em}{0.5em}}
\input{tcilatex}

\begin{document}

\title{Cheating with (Recursive) Models\thanks{%
Financial support from ERC Advanced Investigator grant no. 692995 is
gratefully acknowledges. Eliaz and Spiegler thank Briq and the Economics
Department at Columbia University for their generous hosptiality while this
paper was written. We also thank Armin Falk, Xiaosheng Mu, Martin Weidner,
seminar audiences and especially Heidi Thysen, for helpful comments.}}
\author{Kfir Eliaz\thanks{%
School of Economics, Tel-Aviv University and David Eccles School of
Business, University of Utah. E-mail: kfire@tauex.tau.ac.il.}, Ran Spiegler%
\thanks{%
School of Economics, Tel Aviv University; Department of Economics, UCL; and
CFM. E-mail: rani@tauex.tau.ac.il.} \ and Yair Weiss\thanks{%
School of Computer Science and Engineering, Hebrew University. E-mail:
yweiss@cs.huji.ac.il.}}
\maketitle

\begin{abstract}
To what extent can agents with misspecified subjective models predict false
correlations? We study an \textquotedblleft analyst\textquotedblright\ who
utilizes models that take the form of a recursive system of linear
regression equations. The analyst fits each equation to minimize the sum of
squared errors against an arbitrarily large sample. We characterize the
maximal pairwise correlation that the analyst can predict given a generic
objective covariance matrix, subject to the constraint that the estimated
model does not distort the mean and variance of individual variables. We
show that as the number of variables in the model grows, the false pairwise
correlation can become arbitrarily close to one, regardless of the true
correlation.\bigskip \bigskip \pagebreak
\end{abstract}

\section{Introduction}

Agents in economic models rely on mental models of their environment for
quantifying correlations between variables, the most important of which
describes the payoff consequences of their own actions. The vast majority of
economic models assume rational expectations - i.e., the agent's subjective
models coincide with the modeler's, such that the agent's belief is
consistent with the true data-generating process. Yet when the agent's
subjective model is misspecified, his predicted correlations may deviate
from the truth. This difficulty is faced not only by agents in economic
models but also by real-life researchers in physical and social sciences,
who make use of statistical models. This paper poses a simple question: How
far off can the correlations predicted by misspecified models get?

To make our question concrete, imagine an analyst\ who wishes to demonstrate
to an audience that two variables, $x$ and $y$, are strongly related. Direct
evidence about the correlation between these variables is hard to come by.
However, the analyst has access to data about the correlation of $x$ and $y$
with other variables. He therefore constructs a $model$ that involves $x$, $%
y $ and a selection of auxiliary variables. He fits this model to a large
sample and uses the estimated model to predict the correlation between $x$
and $y$. The analyst is unable (or unwilling) to tamper with the data.
However, he is free to choose the auxiliary variables and how they operate
in the model. To what extent does this degree of freedom enable the analyst
to attain his underlying objective?

This hypothetical scenario is inspired by a number of real-life situations.
First, academic researchers often serve as consultants to policy makers or
activist groups in pursuit of a particular agenda.\ E.g., consider an
economist consulting a policy maker who pursues a tax-cutting agenda and
seeks intellectual support for this position. The policy maker would
therefore benefit from an academic study showing a strong quantitative
relation between tax cuts and economic growth. Second, the analyst may be
wedded to a particular stand regarding the relation between $x$ and $y$
because he staked his public reputation on this claim in the past. Finally,
the analyst may want to make a splash with a counterintuitive finding and
will stop exploring alternative model specifications once he obtains such a
result.

We restrict the class of models that the analyst can employ to be \textit{%
recursive linear-regression models}. A model in this familiar class consists
of a list of linear-regression equations, such that an explanatory variable
in one equation cannot appear as a dependent variable in another equation
down the list. We assume that the recursive model includes the variables $x$
and $y$, as well as a selection of up to $n-2$ additional variables. Thus,
the total number of variables in the analyst's model is $n$, which is a
natural measure of the model's complexity. Each equation is estimated via
Ordinary Least Squares (OLS) against an arbitrarily large (and unbiased)
sample.

The following quote lucidly summarizes two attractions of recursive
models:\medskip

\begin{quote}
\textquotedblleft A system of equations is recursive rather than
simultaneous if there is unidirectional dependency among the endogenous
variables such that, for given values of exogenous variables, values for the
endogenous variables can be determined sequentially rather than jointly. Due
to the ease with which they can often be estimated and the temptation to
interpret them in terms of causal chains, recursive systems were the
earliest equation systems to be used in empirical work in the social
sciences.\textquotedblright \footnote{%
The quote is taken from the International Encyclopedia of the Social
Sciences,
https://www.encyclopedia.com/social-sciences/applied-and-social-sciences-magazines/recursive-models.%
}\medskip
\end{quote}

\noindent The causal interpretation of recursive models is particularly
resonant. If $x$ appears exclusively as an explanatory variable in the
system of equations while $y$ appears exclusively as a dependent variable,
the recursive model intuitively charts a causal explanation that pits $x$ as
a primary cause of $y$, such that the estimated correlation between $x$ and $%
y$ can be legitimately interpreted as an estimated $causal$ effect of $x$ on 
$y$.\pagebreak

\noindent \textit{A three-variable example}

\noindent To illustrate our exercise, suppose that the analyst estimates the
following three-variable recursive model:%
\begin{eqnarray}
x_{1} &=&\varepsilon _{1}  \label{recursivemodel_n=3} \\
x_{2} &=&\beta _{1}x_{1}+\varepsilon _{2}  \notag \\
x_{3} &=&\beta _{2}x_{2}+\varepsilon _{3}  \notag
\end{eqnarray}%
where $x_{1},x_{2},x_{3}$ all have zero mean and unit variance. The analyst
assumes that all the $\varepsilon _{k}$'s are mutually uncorrelated, and
also that for every $k>1$ and $j<k$, $\varepsilon _{k}$ is uncorrelated with 
$x_{j}$ (for $j=k-1$, this is mechanically implied by the OLS method).

For a real-life situation behind this example, consider a pharmaceutical
company that introduces a new drug, and therefore has a vested interest in
demonstrating a large correlation between the dosage of its active
ingredient ($x_{1}$) and the ten-year survival rate associated with some
disease. This correlation cannot be directly measured in the short run.
However, past experience reveals the correlations between the ten-year
survival rate and the levels of various bio-markers (which can serve as the
intermediate variable $x_{2}$). The correlation between these markers and
the drug dosage can be measured experimentally in the short run. Thus, on
one hand the situation calls for a model in the manner of (\ref%
{recursivemodel_n=3}), yet on the other hand the pharmaceutical company's
R\&D unit may select the bio-marker $x_{2}$ opportunistically, in order to
get a large estimated effect. Of course, in reality there may be various
checks and balances that will constrain this opportunism. However, it is
interesting to know how badly it can get, in order to evaluate the
importance for these checks and balances.

Let $\rho _{ij}$ denote the correlation between $x_{i}$ and $x_{j}$
according to the $true$ data-generating process. Suppose that $x_{1}$ and $%
x_{3}$ are objectively uncorrelated - i.e. $r=\rho _{13}=0$. The estimated
correlation between these variables according to the model, given the
analyst's procedure and its underlying assumptions, is%
\begin{equation*}
\hat{\rho}_{13}=\rho _{12}\cdot \rho _{23}
\end{equation*}%
It is easy to see from this expression how the model can generate spurious
estimated correlation between $x_{1}$ and $x_{3}$, even though none exists
in reality. All the analyst has to do is select a variable $x_{2}$ that is
positively correlated with both $x_{1}$ and $x_{3}$, such that $\rho
_{12}\cdot \rho _{23}>0$.

But how large can the false estimated correlation $\hat{\rho}_{13}$ be?
Intuitively, since $x_{1}$ and $x_{3}$ are objectively uncorrelated, if we
choose $x_{2}$ such that it is highly correlated with $x_{1}$, its
correlation with $x_{3}$ will be low. In other words, increasing $\rho _{12}$
will come at the expense of decreasing $\rho _{23}$. Formally, consider the
true correlation matrix:%
\begin{equation*}
\begin{array}{ccc}
1 & \rho _{12} & 0 \\ 
\rho _{12} & 1 & \rho _{23} \\ 
0 & \rho _{23} & 1%
\end{array}%
\end{equation*}%
By definition, this matrix is positive semi-definite. This property is
characterized by the inequality $(\rho _{12})^{2}+(\rho _{23})^{2}\leq 1$.
The maximal value of $\rho _{12}\cdot \rho _{23}$ subject to this constraint
is $\frac{1}{2}$, and therefore this is the maximal false correlation that
the above recursive model can generate. This bound is tight: It can be
attained if we define $x_{2}$ to be a deterministic function of $x_{1}$ and $%
x_{3}$, given by $x_{2}=\frac{1}{2}(x_{1}+x_{3})$. Thus, while a given
misspecified recursive model may be able to generate spurious estimated
correlation between objectively independent variables, there is a limit to
how far it can go.

Our interest in the upper bound on $\hat{\rho}_{13}$ is not purely
mathematical. As the above \textquotedblleft bio-marker\textquotedblright\
example implies, we have in mind situations in which the analyst can select $%
x_{2}$ from a \textit{large pool} of potential auxiliary variable. In the
current age of \textquotedblleft big data\textquotedblright , analysts have
access to datasets involving a huge number of covariates. As a result, they
have considerable freedom when deciding which variables to incorporate into
their models. This helps our analyst generate a false correlation that
approaches the theoretical upper bound.

\begin{figure}[tbp]
\centerline{\includegraphics[width=0.46\columnwidth]{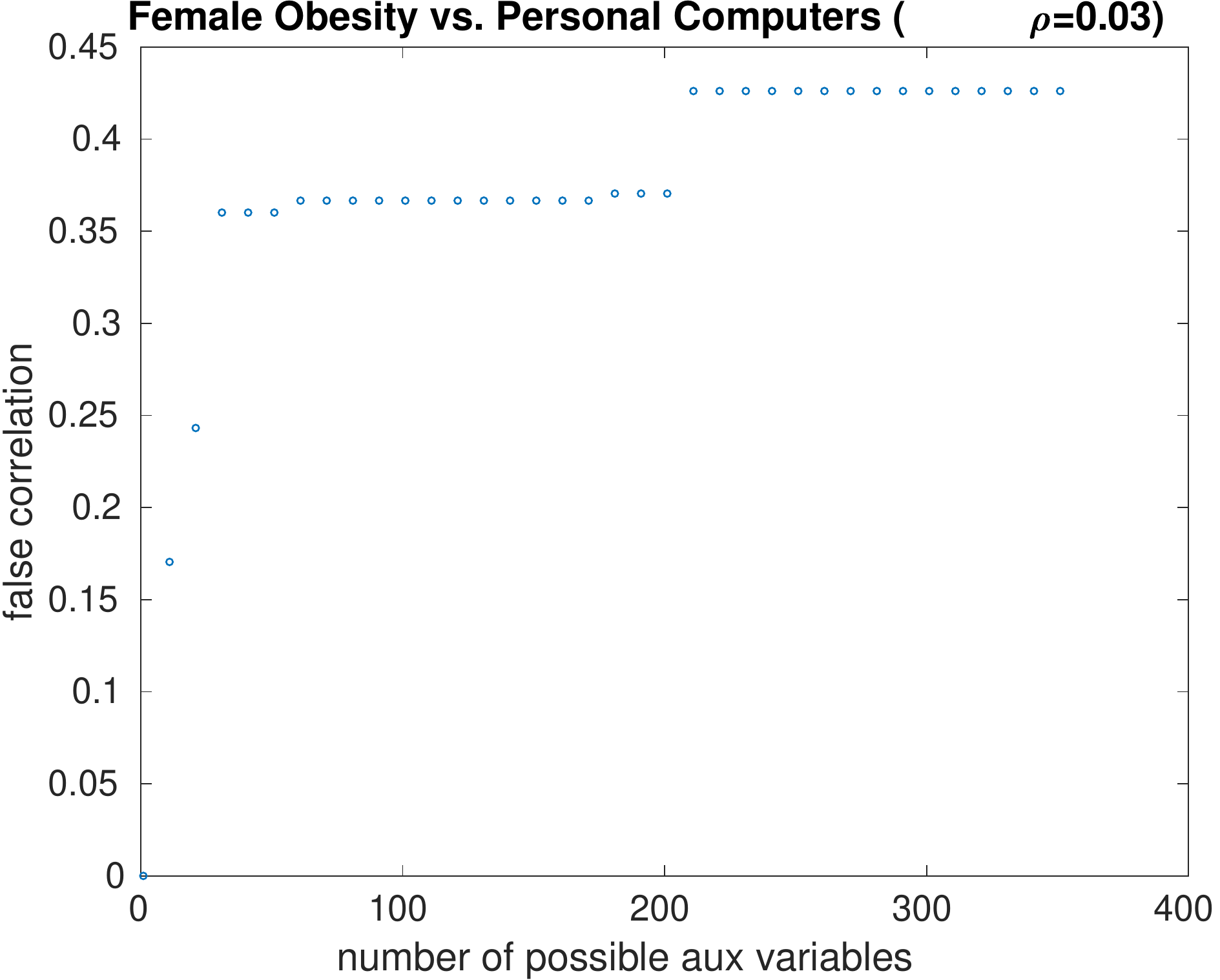}
    \includegraphics[width=0.5\columnwidth]{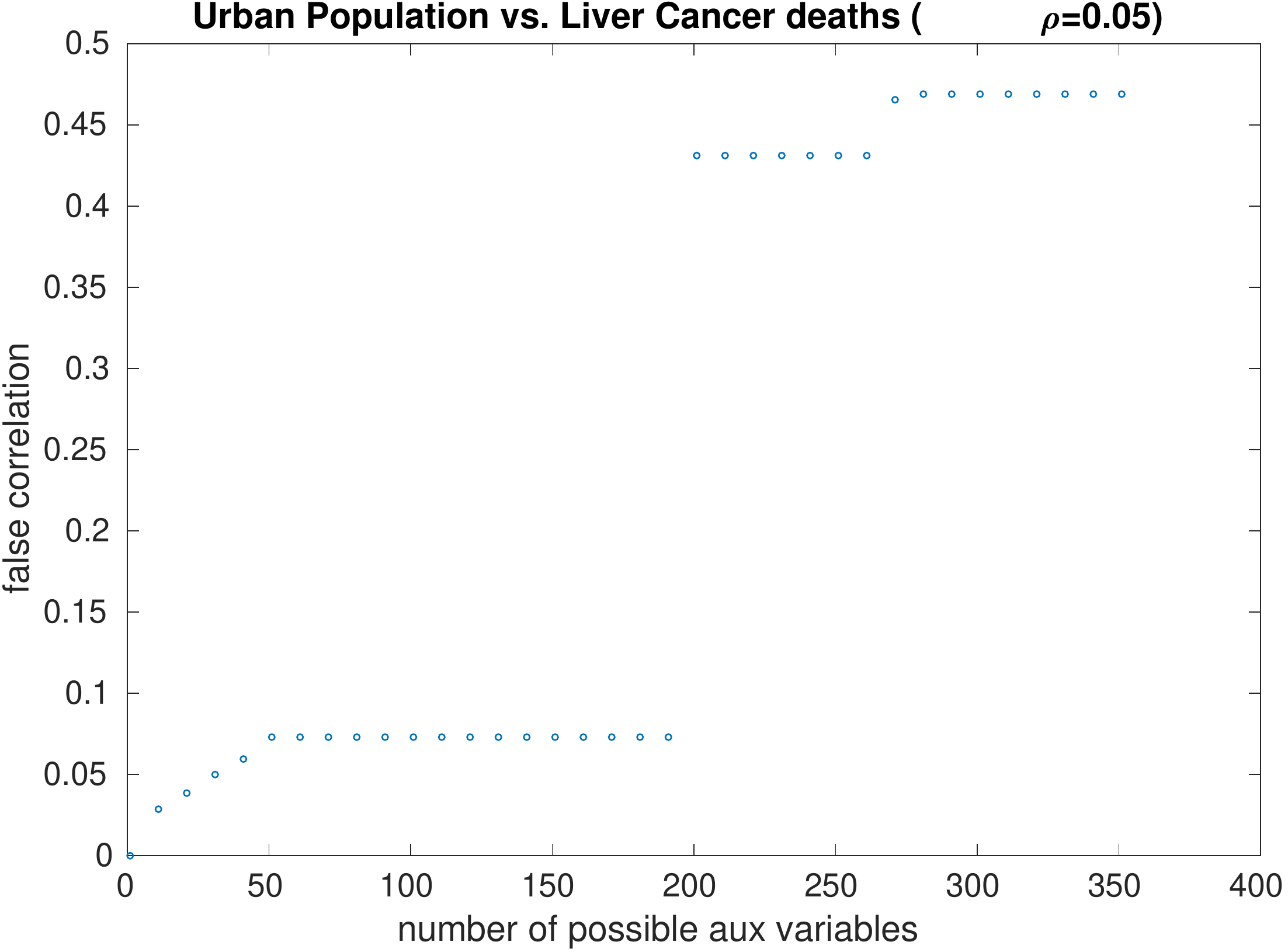} 
  }
\caption{False correlation in a recursive model with one auxiliary variable,
as a function of the number of possible auxiliary variables the researcher
can choose from. All variables and their correlations are taken from a
database compiled by the World Health Organization. Even though the true
correlation is close to zero in both cases, as the number of possible
auxilary variable increases, the estimated correlation rises yet never
exceeds $0.5$.}
\end{figure}

To give a concrete demonstration for this claim, consider Figure $1$, which
is extracted from a database compiled by the World Health Organization and
collected by Reshef et al. (2011).\footnote{%
The variables are collected on all countries in the WHO database (see
www.who.int/whosis/en/) for the year 2009.} All the variables are taken from
this database. The figure displays the maximal $\hat{\rho}_{13}$ correlation
that the model (\ref{recursivemodel_n=3}) can generate for two fixed pairs
of variables $x_{1}$ and $x_{3}$ with $\rho _{13}$ close to zero, when the
auxiliary variable is selected from a pool whose size is given by the
horizontal axis (variables were added to the pool in some arbitrary order).
When the analyst can choose $x_{2}$ from only ten possible auxiliary
variables, the estimated correlation between $x_{1}$ and $x_{3}$ he can
generate with (\ref{recursivemodel_n=3}) is still modest. In contrast, once
the pool size is in the hundreds, the estimated correlation approaches the
upper bound of $\frac{1}{2}$.

For a specific variable that gets us near the theoretical upper bound,
consider the figure's R.H.S, where $x_{1}$ represents urban population and $%
x_{3}$ represents liver cancer deaths per 100,000 men. The true correlation
between these variables is $0.05$. If the analyst selects $x_{2}$ to be coal
consumption (measured in tonnes oil equivalent), the estimated correlation
between $x_{1}$ and $x_{3}$ is $0.43$, far above the objective value. This
selection of $x_{2}$ has the added advantage that the model suggests an
intuitive causal mechanism: Urbanization causes cancer deaths via its effect
on coal consumption.\medskip

\noindent \textit{Review of the results}

\noindent We present our formal model in Section 2 and pose our main
problem: What is the largest estimated correlation between $x_{1}$ and $%
x_{n} $ that a recursive, $n$-variable linear-regression model can generate?
We impose one constraint on this maximization problem: While the estimated
model is allowed to distort correlations among variables, it must produce
correct estimates of the individual variables' mean and variance. The
linearity of the analyst's model implies that only the latter has bite.

To motivate this constraint, recall that the scenario behind our model
involves a sophisticated analyst and a lay audience. Because the audience is
relatively unsophisticated, it cannot be expected to discipline the
analyst's opportunistic model selection with elaborate tests for model
misspecification that involve conditional or unconditional correlations.
However, monitoring \textit{individual} variables is a much simpler task
than monitoring correlations between variables. E.g., it is relatively easy
to disqualify an economic model that predicts highly volatile inflation if
observed inflation is relatively stable. Likewise, a climatological model
that underpredicts temperature volatility loses credibility, even for a lay
audience.

Beyond this justification, we simply find it intrinsically interesting to
know the extent to which misspecified recursive models can distort
correlations between variables while preserving moments of individual
variables. At any rate, we relax the constraint in Section 4, for the
special case of models that consist of a single non-degenerate regression
equation. We use this case to shed light on the analyst's opportunistic use
of \textquotedblleft \textit{bad controls}\textquotedblright .

In Section 3, we derive the following result. For a generic objective
covariance matrix with $\rho _{1n}=r$, the maximal estimated correlation $%
\hat{\rho}_{1n}$ that a recursive model with up to $n$ variables can
generate, subject to preserving the mean and variance of individual
variables, is%
\begin{equation}
\left( \cos \left( \frac{\arccos r}{n-1}\right) \right) ^{n-1}
\label{formula}
\end{equation}%
The upper bound given by (\ref{formula}) is tight. Specifically, it is
attained by the simplest recursive model that involves $n$ variables: For
every $k=2,...,n$, $x_{k}$ is regressed on $x_{k-1}$ only. This model is
represented graphically by the chain $x_{1}\rightarrow x_{2}\rightarrow
\cdots \rightarrow x_{n}$. his chain has an intuitive causal interpretation,
which enables the analyst to present $\hat{\rho}_{1n}$ as an estimated
causal effect of $x_{1}$ on $x_{n}$. The variables $x_{2},...,x_{n-1}$ that
are employed in this model have a simple definition, too: They are all
deterministic linear functions of $x_{1}$ and $x_{n}$.

Formula (\ref{formula}) reproduces the value $\hat{\rho}_{13}=\frac{1}{2}$
that we derived in our illustrative example, and it is strictly increasing
in $n$. When $n\rightarrow \infty $, the expression converges to $1$. That
is, regardless of the true correlation between $x_{1}$ and $x_{n}$, a
sufficiently large recursive model can generate an arbitrarily large
estimated correlation. The lesson is that when the analyst is free to select
his model and the variables that inhabit it, he can deliver any conclusion
about the effect of one variable on another - unless we impose constraints
on his procedure, such as bounds on the complexity of his model or
additional misspecification tests.

The formula (\ref{formula}) has a simple geometric interpretation, which
also betrays the construction of the recursive model and objective
covariance matrix that implement the upper bound. Take the angle that
represents the objective correlation $r$ between $x_{1}$ and $x_{n}$; divide
it into $n-1$ equal sub-angles; this sub-angle represents the correlation
between adjacent variables along the above causal chain; the product of
these correlations produces the estimated correlation between $x_{1}$ and $%
x_{n}$.

The detailed proof of our main result is presented in\ Section 5. It relies
on the graphical representation of recursive models and employs tools from
the Bayesian-networks literature (Cowell et al. (1999), Koller and Friedman
(2009)). In the Appendix, we present partial analysis of our question for a
different class of models involving $binary$ variables.\medskip

\noindent \textit{Related literature}

\noindent There is a huge literature on misspecified models in various
branches of economics and statistics, which is too vast to survey in detail
here. A few recent references can serve as entry points for the interested
reader: Esponda and Pouzo (2016), Bonhomme and Weidner (2018) and Molavi
(2019). To our knowledge, our paper is the first to carry out a worst-case
analysis of misspecified models' predicted correlations.

The \textquotedblleft analyst\textquotedblright\ story that motivated this
exercise brings to mind the phenomenon of researcher bias. A few works in
Economics have explicitly modeled this bias and its implications for
statistical inference. (Of course, there is a larger literature on how
econometricians should cope with researcher/publication bias, but here we
only describe exercises that contain explicit models of the researcher's
behavior.) Leamer (1974) suggests a method of discounting evidence when
linear regression models are constructed after some data have been partially
analyzed. Lovell (1983) considers a researcher who chooses $k$ out of $n$
independent variables as explanatory variables in a single regression with
the aim of maximizing the coefficient of correlation between the chosen
variables and the dependent variable. He argues that a regression
coefficient that appears to be significant at the $\alpha $ level should be
regarded as significant at only the $1-(1-\alpha )^{n/k}$ level. Glaeser
(2008) suggests a way of correcting for this form of data mining in the
coefficient estimate.

More recently, Di-Tillio, Ottaviani and Sorensen (2017,2019) characterize
data distributions for which strategic sample selection (e.g., selecting the 
$k$ highest observations out of $n$) benefits an evaluator who must take an
action after observing the selected sample realizations. Finally, Spiess
(2018) proposes a mechanism-design framework to align the preferences of the
researcher with that of \textquotedblleft society\textquotedblright :\ A
social planner first chooses a menu of possible\textit{\ estimators}, the
investigator chooses an estimator from this set, and the estimator is then
applied to the sampled observations.

The analyst in our model\ need not be a scientist - he could also be a
politician or a pundit. Under this interpretation, constructing a model and
fitting it to data is not an explicit, formal affair. Rather, it involves
spinning a \textquotedblleft narrative\textquotedblright\ about the effect
of policy on consequences and using casual empirical evidence to
substantiate it. Eliaz and Spiegler (2018) propose a model of political
beliefs that is based on this idea. From this point of view, our exercise in
this paper explores the extent to which false narratives can exaggerate the
effect of policy.

\section{The Model}

Let $p$ be an objective probability measure over $n$ variables, $%
x_{1},...,x_{n}$. For every $A\subset \{1,...,n\}$, denote $%
x_{A}=(x_{i})_{i\in A}$. Assume that the marginal of $p$ on each of these
variables has\textit{\ zero mean and unit variance.} This will entail no
loss of generality for our purposes. We use $\rho _{ij}$ to denote the
coefficient of correlation between the variables $x_{i},x_{j}$, according to 
$p$. In particular, denote $\rho _{1n}=r$. The covariance matrix that
characterizes $p$ is therefore $(\rho _{ij})$.

An analyst estimates a recursive model that involves these variables. This
model consists of a system of linear-regression equations. For every $%
k=1,...,n$, the $k^{th}$ equation takes the form%
\begin{equation*}
x_{k}=\mathop{\displaystyle \sum }\limits_{j\in R(k)}\beta
_{jk}x_{j}+\varepsilon _{k}
\end{equation*}%
where:\medskip

\begin{itemize}
\item $R(k)\subseteq \{1,..,k-1\}$. This restriction captures the model's
recursive structure: An explanatory variable in one equation cannot appear
as a dependent variable in a later equation.

\item In the $k^{th}$ equation, the $\beta _{jk}$'s are parameters to be
estimated against an infinitely large sample drawn from $p$. The analyst
assumes that each $\varepsilon _{k}$ has zero mean and that it is
uncorrelated with all other $\varepsilon _{j}$'s, as well as with $%
(x_{j})_{j<k}$. The $\beta _{jk}$'s are selected to minimize the mean
squared error of the $k^{th}$ regression equation, which gives the standard
Ordinary Least Squares estimate:%
\begin{equation}
\beta _{k}=\rho _{k,R(k)}\left( \rho _{R(k),R(k)}\right) ^{-1}
\label{beta_i}
\end{equation}%
where $\beta _{k}=(\beta _{jk})_{j\in R(k)}$, $\rho _{k,R(k)}$ denotes the
row of correlations between $x_{k}$ and each of the explanatory variables $%
x_{j}$, $j\in R(k)$, and $\rho _{R(k),R(k)}$ denotes the submatrix of the
correlations among the explanatory variables.\medskip
\end{itemize}

We refer to such a system of regression equations as an $n$\textit{-variable
recursive model}. The function $R$ effectively defines a directed acyclic
graph (DAG) over the set of nodes $\{1,...,n\}$, such that a link $%
i\rightarrow j$ exists whenever $i\in R(j)$. DAGs are often interpreted as
causal models (see Pearl (2009)). We will make use of the DAG representation
in the proof of our main result, as well as in the Appendix. We will also
allude to the recursive model's causal interpretation, but without
addressing explicitly the question of causal inference. The analyst in our
model engages in a problem of fitting a model to data. While the causal
interpretation may help the analyst to \textquotedblleft
sell\textquotedblright\ the model to his audience, he does not engage in an
explicit procedure for drawing causal inference from his model.

Note that in the analyst's model, the equation for $x_{1}$ has no
explanatory variables, and $x_{n}$ is not an explanatory variable in any
equation. Furthermore, the partial ordering given by $R$ is consistent with
the natural enumeration of variables (i.e., $i\in R(j)$ implies $i<j$). This
restriction is made for notational convenience; relaxing it would not change
our results. However, it has the additional advantage that the causal
interpretation of the model-estimated correlation between $x_{1}$ and $x_{n}$
is sensible. Indeed, it is legitimate according to Pearl's (2009) rules for
causal inference based on DAG-represented models.

The analyst's assumption that each $\varepsilon _{k}$ is uncorrelated with $%
x_{R(k)}$ is redundant, because it is an automatic consequence of his OLS
procedure for estimating $\beta _{k}$. In contrast, his assumption that $%
\varepsilon _{k}$ is uncorrelated with \textit{all other} $x_{j}$, $j<k$, is
the basis for how he combines the individually estimated equations into a
joint estimated distribution. It is fundamentally a conditional-independence
assumption - namely, that $x_{k}$ is independent of $(x_{j})_{j\in
\{1,..,k-1\}-R(k)}$ conditional on $x_{R(k)}$. This assumption will
typically be false - indeed, it is what makes his model misspecified and
what enables him to \textquotedblleft cheat\textquotedblright\ with his
model.

Under this assumption, the analyst proceeds to estimate the correlation
between $x_{1}$ and $x_{n}$, which can be computed according to the
following recursive procedure. Start with the $n^{th}$ equation. For every $%
j\in R(n)-\{1\}$, replace $x_{j}$ with the R.H.S of the $j^{th}$ equation.
This produces a new equation for $x_{n}$, with a different set of
explanatory variables. Repeat the substitution for each one of these
variables, and continue doing so until the only remaining explanatory
variable is $x_{1}$.

The procedure's final output is the thus the equation%
\begin{equation*}
x_{n}=\alpha x_{1}+\sum_{j=1}^{n}\gamma _{j}\varepsilon _{j}
\end{equation*}%
The coefficients $\alpha ,\gamma _{1},...,\gamma _{n}$ are combinations of $%
\beta $ parameters (which were obtained by OLS estimation of the individual
equations). Likewise, the distribution of each error term $\varepsilon _{j}$
is taken from the estimated $j^{th}$ equation.

The analyst uses this equation to estimate the variance of $x_{n}$ and its
covariance with $x_{1}$, implementing the (partly erroneous) assumptions
that all the $x_{k}$'s have zero mean and unit variance, and that the $%
\varepsilon _{k}$'s mean, mutual covariance and covariance with $x_{1}$ are
all zero:%
\begin{eqnarray*}
\widehat{Var}(x_{n}) &=&\alpha ^{2}\cdot Var(x_{1})+\sum_{j=1}^{n}(\gamma
_{j})^{2}Var(\varepsilon _{j}) \\
&=&\alpha ^{2}+\sum_{j=1}^{n}(\gamma _{j})^{2}Var(\varepsilon _{j})
\end{eqnarray*}%
and%
\begin{eqnarray*}
\widehat{Cov}(x_{1},x_{n}) &=&E\left[ x_{1}\left( \alpha
x_{1}+\sum_{j=1}^{n}\gamma _{j}\varepsilon _{j}\right) \right] \\
&=&\alpha \cdot Var(x_{1})+\sum_{j=1}^{n}\gamma _{j}E(x_{1}\varepsilon _{j})
\\
&=&\alpha
\end{eqnarray*}%
Therefore, the estimated coefficient of correlation between $x_{1}$ and $%
x_{n}$ is%
\begin{equation}
\hat{\rho}_{1n}=\frac{\widehat{Cov}(x_{1},x_{n})}{\sqrt{Var(x_{1})\widehat{%
Var}(x_{n})}}=\frac{\alpha }{\sqrt{\widehat{Var}(x_{n})}}
\label{estimated_correlation}
\end{equation}

We assume that the analyst faces the constraint that the estimated mean and
variance of all individual variables must be correct (see the Introduction
for a discussion of this constraint). The requirement that the estimated
means of individual variables are undistorted has no bite: The OLS procedure
for individual equations satisfies it. Thus, the constraint is reduced to
the requirement that%
\begin{equation*}
\widehat{Var}(x_{k})=1
\end{equation*}%
for all $k$ (we can calculate this estimated variance for every $k>1$, using
the same recursive procedure we applied to $x_{n}$). This reduces (\ref%
{estimated_correlation}) to%
\begin{equation*}
\hat{\rho}_{1n}=\alpha
\end{equation*}%
Our objective will be to examine how large this expression can be, given $n$
and a generic objective covariance matrix problem.\medskip

\noindent \textit{Comments on the analyst's procedure}

\noindent In our model, the analyst relies on a structural model to generate
an estimate of the correlation between $x_{1}$ and $x_{n}$, which he
presents to a lay audience. The process by which he selects\ the model
remains hidden from the audience. But why does the analyst use a model to
estimate the correlation between $x_{1}$ and $x_{n}$, rather than estimating
it $directly$? One answer may be that direct evidence on this correlation is
hard to come by (as, for example, in the case of long-term health effects of
nutritional choices). In this case, the analyst $must$ use a model to
extrapolate an estimate of $\rho _{1n}$ from observed data.

Another answer is that analysts use models as simplified representations of
a complex reality, which they can consult for $multiple$
conditional-estimation tasks: Estimating the effect of $x_{1}$ on $x_{n}$ is
only one of these tasks. This is illustrated by the following quote:
\textquotedblleft The economy is an extremely complicated mechanism, and
every macroeconomic model is a vast simplification of reality\ldots the
large scale of FRB/US [\textit{a general equilibrium model employed by the
Federal Reserve Bank - the authors}] is an advantage in that it can perform
a wide variety of computational `what if' experiments.\textquotedblright 
\footnote{%
This quote is taken from a speech by Stanley Fisher: See
https://www.federalreserve.gov/newsevents/speech/fischer20170211a.htm.} From
this point of view, our analysis concerns the maximal distortion of pairwise
correlations that such models can produce.

Our treatment of the model accommodates both interpretations. In particular,
we will allow for the possibility that $1\in R(n)$, which means that the
analyst does have data about the joint distribution of $x_{1}$ and $x_{n}$.
As we will see, our main result will not make use of this possibility.

\section{The Main Result}

For every $r,n$, denote%
\begin{equation*}
\theta _{r,n}=\frac{\arccos r}{n-1}
\end{equation*}%
We are now able to state our main result.\medskip

\begin{theorem}
\label{thm1} For almost every true covariance matrix $(\rho _{ij})$
satisfying $\rho _{1n}=r$, if the estimated recursive model satisfies $%
\widehat{Var}(x_{k})=1$ for all $k$, then the estimated correlation between $%
x_{1}$ and $x_{n}$ satisfies%
\begin{equation*}
\hat{\rho}_{1n}\leq \left( \cos \theta _{r,n}\right) ^{n-1}
\end{equation*}%
$\medskip $Moreover, this upper bound can be implemented by the following
pair:\medskip \newline
(i) A recursive model defined by $R(k)=\{k-1\}$ for every $k=2,...,n$.%
\newline
(ii) A multivariate Gaussian distribution satisfying, for every $k=1,...,n$:%
\begin{equation}
x_{k}=s_{1}\cos ((k-1)\theta _{r,n})+s_{2}\sin ((k-1)\theta _{r,n})\medskip
\label{solution_xk}
\end{equation}%
where $s_{1},s_{2}$ are independent standard normal variables.\medskip
\end{theorem}

Let us illustrate the upper bound given by Theorem \ref{thm1} numerically
for the case of $r=0$, as a function of $n$:%
\begin{equation*}
\begin{array}{ccccc}
n & 2 & 3 & 4 & 5 \\ 
\text{upper bound on }\hat{\rho}_{1n} & 0 & 0.5 & 0.65 & 0.73%
\end{array}%
\end{equation*}%
As we can see, the marginal contribution of adding a variable to the false
correlation that the analyst's model can produce decays quickly. However,
when $n\rightarrow \infty $, the upper bound converges to one. This is the
case for any value of $r$. That is, even if the true correlation between $%
x_{1}$ and $x_{n}$ is strongly negative, a sufficiently large model can
produce a large positive correlation.

The recursive model that attains the upper bound has a simple structure. Its
DAG representation is a single chain%
\begin{equation*}
1\rightarrow 2\rightarrow \cdots \rightarrow n
\end{equation*}%
Intuitively, this is the simplest connected DAG with $n$ nodes: It has the
smallest number of links among this class of DAGs, and it has no junctions.
The distribution over the auxiliary variables $x_{2},...,x_{n}$ in the upper
bound's implementation has a simple structure, too: Every $x_{k}$ is a
different linear combination of two independent \textquotedblleft
factors\textquotedblright , $s_{1}$ and $s_{2}$. We can identify $s_{1}$
with $x_{1}$, without loss of generality. The closer the variable lies to $%
x_{1}$ along the chain, the larger the weight it puts on $s_{1}$.$\medskip $

\noindent \textit{General outline of the proof}

\noindent The proof of Theorem \ref{thm1} proceeds in three major steps.
First, the constraint that the estimated model preserves the variance of
individual variables for a generic objective distribution reduces the class
of candidate recursive models to those that can be represented by \textit{%
perfect} DAGs. Since perfect DAGs preserve marginals of individual variables
for $every$ objective distribution (see Spiegler (2017)), the theorem can be
stated more strongly for this subclass of recursive models.\medskip

\begin{proposition}
\label{prop_perfect}Consider a recursive model given by $R$. Suppose that
for every $k>1$, if $i,j\in R(k)$ and $i<j$, then $i\in R(j)$. Then, for
every true covariance matrix $(\rho _{ij})$ satisfying $\rho _{1n}=r$, $\hat{%
\rho}_{1n}\leq \left( \cos \theta _{r,n}\right) ^{n-1}$.\medskip
\end{proposition}

\noindent That is, when the recursive model is represented by a perfect DAG,
the upper bound on $\hat{\rho}_{1n}$ holds for $any$ objective covariance
matrix, and the undistorted-variance constraint is redundant.

In the second step, we use the tool of \textit{junction trees} in the
Bayesian-networks literature (Cowell et al. (1999)) to perform a further
reduction in the class of relevant recursive models. Consider a recursive
model represented by a non-chain perfect DAG. We show that the analyst can
generate the same $\hat{\rho}_{1n}$ with another objective distribution and
a recursive model that takes the form of a simple chain $1\rightarrow \cdots
\rightarrow n$. Furthermore, this chain will involve no more variables than
the original model.

To illustrate this argument, consider the following recursive model with $%
n=4 $:%
\begin{eqnarray*}
x_{1} &=&\varepsilon _{1} \\
x_{2} &=&\beta _{12}x_{1}+\varepsilon _{2} \\
x_{3} &=&\beta _{13}x_{1}+\beta _{23}x_{2}+\varepsilon _{3} \\
x_{4} &=&\beta _{24}x_{2}+\beta _{34}x_{3}+\varepsilon _{4}
\end{eqnarray*}%
This recursive model has the following DAG representation:%
\begin{equation*}
\begin{array}{ccc}
1 & \rightarrow & 3 \\ 
\downarrow & \nearrow & \downarrow \\ 
2 & \rightarrow & 4%
\end{array}%
\end{equation*}%
Because $x_{4}$ depends on $x_{2}$ and $x_{3}$ only through their linear
combination $\beta _{24}x_{2}+\beta _{34}x_{3}$, we can replace $%
(x_{2},x_{3})$ with a scalar variable $x_{5}$, such that the recursive model
becomes%
\begin{eqnarray*}
x_{1} &=&\varepsilon _{1} \\
x_{5} &=&\beta _{15}^{\prime }x_{1}+\varepsilon _{5}^{\prime } \\
x_{4} &=&\beta _{54}^{\prime }x_{5}+\varepsilon _{4}
\end{eqnarray*}%
This model is represented by the DAG $1\rightarrow 5\rightarrow 3$, which is
a simple chain that consists of fewer nodes than the original DAG.

This means that in order to calculate the upper bound on $\hat{\rho}_{1n}$,
we can restrict attention to the chain model. But in this case, the
analyst's objective function has a simple explicit form:%
\begin{equation*}
\hat{\rho}_{1n}=\mathop{\displaystyle \prod }\limits_{k=1}^{n-1}\rho _{k,k+1}
\end{equation*}%
Thus, in the third step, we derive the upper bound by finding the
correlation matrix that maximizes the R.H.S of this formula, subject to the
constraints that $\rho _{1n}=r$ and that the matrix is positive
semi-definite (which is the property that defines the class of covariance
matrices). The solution to this problem has a simple geometric
interpretation.

\section{Single-Equation Models}

Analysts often propose models that take the form of a $single$
linear-regression equation, consisting of a dependent variable $x_{n}$
(where $n>2$), an explanatory variable $x_{1}$ and $n-2$ \textquotedblleft $%
control$\textquotedblright\ variables $x_{2},...,x_{n-1}$ . Using the
language of Section 2, this corresponds to the specification $R(k)=\emptyset 
$ for all $k=1,...,n-1$ and $R(n)=\{1,...,n-1\}$. That is, the only
non-degenerate equation is the one for $x_{n}$, hence the term
\textquotedblleft single-equation model\textquotedblright . Note that in
this case, the OLS regression coefficient $\beta _{1n}$ in the equation for $%
x_{n}$ coincides with $\hat{\rho}_{1n}=\alpha /\widehat{Var}(x_{n})$, as
defined in Section 2.

Using the graphical representation, the single-equation model corresponds to
a DAG in which $x_{1},...,x_{n-1}$ are all ancestral nodes that send links
into $x_{n}$. Since this DAG is imperfect, Lemma \ref{lemma1} in Section 5
implies that for almost all objective covariance matrices, the estimated
variance of $x_{n}$ according to the single-equation model will differ from
its true value.\footnote{%
All the other variables are represented by ancestral nodes, and therefore
their marginals are not distorted (see Spiegler (2017)).} However, given the
particular interest in this class of models, we relax the correct-variance
constraint in this section and look for the maximal false correlation that
such models can generate. For expositional convenience, we focus on the case
of $r=0$.\medskip

\begin{proposition}
\label{prop}Let $r=0$. Then, a single-equation model $x_{n}=\sum_{i=1}^{n-1}%
\beta _{i}x_{i}+\varepsilon $ can generate an estimated coefficient $\hat{%
\rho}_{1,n}$ of at most $1/\sqrt{2}$. This bound is tight, and can be
approximated arbitrarily well with $n=3$ such that $x_{2}=\delta x_{1}+\sqrt{%
1-\delta ^{2}}x_{3}$, where $\delta \approx -1$.
\end{proposition}

\begin{proof}
Because $x_{2},...,x_{n-1}$ are Gaussian without loss of generality, we can
replace their linear combination $(\sum_{i=2}^{n-1}\beta
_{i}x_{i})/(\sum_{i=2}^{n-1}\beta _{i})$ (where the $\beta _{i}$'s are
determined by the objective $p$) by a single Gaussian variable $z$ that has
mean zero, but its variance need not be one. Its objective distribution
conditional on $x_{1},x_{n}$ can be written as a linear equation $z=\delta
x_{1}+\gamma x_{n}+\eta $. Since all variables on the R.H.S of this equation
are independent (and since $x_{1}$ and $x_{n}$ are standardized normal
variables), it follows that the objective variance of $z$ is%
\begin{equation*}
Var(z)=\delta ^{2}+\gamma ^{2}+\sigma ^{2}
\end{equation*}

The analyst's model can now be written as%
\begin{equation}
x_{n}=\frac{1}{\gamma }z-\frac{\delta }{\gamma }x_{1}-\frac{1}{\gamma }\eta
\label{singleequation}
\end{equation}%
Our objective is to find the values of $\delta $, $\gamma $ and $\sigma $
that maximize 
\begin{equation*}
\hat{\rho}_{1,n}=\frac{\hat{E}(x_{1},x_{n})}{\sqrt{\widehat{Var}(x_{n})%
\widehat{Var}(x_{1})}}
\end{equation*}%
Because $x_{1}$ and $x_{n}$ are independent, standardized normal, $\hat{E}%
(x_{1},x_{n})=-\delta /\gamma $. The analyst's model does not distort the
variance of $x_{1}$.\footnote{%
The reason is that the node that represents $x_{1}$ in the DAG
representation of the model is ancestral. By Spiegler (2017), the estimated
model does not distort the marginals of such variables.} Therefore, $%
\widehat{Var}(x_{1})$. And since the analyst's model regards $z$, $x_{1}$
and $\eta $ as independent,%
\begin{equation*}
\widehat{Var}(x_{n})=\left( \frac{1}{\gamma }\right) ^{2}Var(z)+\left( \frac{%
\delta }{\gamma }\right) ^{2}+\left( \frac{\sigma }{\gamma }\right)
^{2}=\left( \frac{1}{\gamma }\right) ^{2}\left( \delta ^{2}+\gamma
^{2}+\sigma ^{2}\right) +\left( \frac{\delta }{\gamma }\right) ^{2}+\left( 
\frac{\sigma }{\gamma }\right) ^{2}
\end{equation*}%
It is clear from this expression that in order to maximize $\hat{\rho}_{1,n}$%
, we should set $\sigma =0$. It follows that%
\begin{equation*}
\hat{\rho}_{1,n}=-\frac{\frac{\delta }{\gamma }}{\sqrt{1+2\left( \frac{%
\delta }{\gamma }\right) ^{2}}}
\end{equation*}%
which is decreasing in $\delta /\gamma $ and attains an upper bound of $1/%
\sqrt{2}$ when $\delta /\gamma \rightarrow -\infty $.

Note that since without loss of generality we can set $\gamma =\sqrt{%
1-\delta ^{2}}$ such that $z\sim N(0,1)$. Therefore, the upper bound is
approximated to an arbitrarily fine degree when we set $\delta \rightarrow
-1 $ $\ $such that $\gamma \rightarrow 0$. As a result, the estimated
variance of $x_{n}$ diverges.\medskip
\end{proof}

Thus, to magnify the false correlation between $x_{1}$ and $x_{n}$, the
analyst would select the \textquotedblleft control\textquotedblright\
variables such that a certain linear combination of them has strong negative
correlation with $x_{1}$. That is, the analyst will prefer his regression
model to exhibit multicollinearity. This inflates the estimated variance of $%
x_{n}$; indeed, $\widehat{Var}(x_{3})\rightarrow \infty $ when $\delta
\rightarrow -1$. However, at the same time it increases the estimated
covariance between $x_{1}$ and $x_{3}$, which more than compensates for this
increase in variance. As a result, the estimated correlation between $x_{1}$
and $x_{3}$ rises substantially.

The three-variable model that implements the upper bound is represented by
the DAG $1\rightarrow 3\leftarrow 2$. That is, it treats the variables $%
x_{1} $ and $x_{2}$ as independent, even though in reality they are
correlated. In particular, the objective distribution may be consistent with
a DAG that adds a link $1\rightarrow 2$ to the analyst's DAG, such that
adding $x_{2}$ to the regression means that we control for a
\textquotedblleft post-treatment\textquotedblright\ variable (where $x_{1}$
is viewed as the treatment). In other words, $x_{2}$ is a \textquotedblleft
bad control\textquotedblright\ (see Angrist and Pischke (2008), p. 64).%
\footnote{%
See also
http://causality.cs.ucla.edu/blog/index.php/2019/08/14/a-crash-course-in-good-and-bad-control/.%
}

The upper bound of $1/\sqrt{2}$ in Proposition \ref{prop} is obtained with $%
n=3$. Recall that under the undistorted-variance constraint, the upper bound
on $\hat{\rho}_{1,n}$ for $n=3$ is $1/2$. This shows that the constraint has
bite. However, when $n$ is sufficiently large, the single-equation model is
outperformed by the multi-equation chain model, which does satisfy the
undistorted-variance constraint.

\section{Proof of Theorem \protect\ref{thm1}}

\subsection{Preliminaries: Bayesian Networks}

The proof relies on concepts and tools from the Bayesian-network literature
(Cowell et al. (1999), Koller and Friedman (2009)). Therefore, we introduce
a few definitions that will serve us in the proof.

A DAG is a pair $G=(N,R)$, where $N$ is a set of nodes and $R\subset N\times
N$ is a pair of directed links. We assume throughout that $N=\{1,...,n\}$.
With some abuse of notation, $R(i)$ is the set of nodes $j$ for which the
DAG includes a link $j\rightarrow i$. A DAG is $perfect$ if whenever $i,j\in
R(k)$ for some $i,j,k\in N$, it is the case that $i\in R(j)$ or $j\in R(i)$.

A subset of nodes $C\subseteq N$ is a $clique$ if for every $i,j\in C$, $iRj$
or $jRi$. We say that a clique is \textit{maximal} if it is not contained in
another clique. We use $\mathcal{C}$ to denote the collection of maximal
cliques in a DAG.

A node $i\in N$ is ancestral if $R(i)$ is empty. A node $i\in N$ is terminal
if there is no $j\in N$ such that $i\in R(j)$. In line with our definition
of recursive models in Section 2, we assume that $1$ is ancestral and $n$ is
terminal. It is also easy to verify that we can restrict attention to DAGs
in which $n$ is the $only$ terminal node - otherwise, we can remove the
other terminal nodes from the DAG, without changing $\hat{p}(x_{n}\mid
x_{1}) $. We will take these restrictions for granted henceforth.

The analyst's procedure for estimating $\hat{\rho}_{1n}$, as described in
Section 2, has an equivalent description in the language of Bayesian
networks, which we now describe.

Because the analyst estimates a linear model, it is \textit{as if} he
believes that the underlying distribution $p$ is \textit{multivariate normal}%
, where the estimated $k^{th}$ equation is a complete description of the
conditional distribution $(p(x_{k}\mid x_{R(k)}))$. Therefore, from now, we
will proceed as if $p$ were indeed a standardized multivariate normal with
covariance matrix $(\rho _{ij})$, such that the $k^{th}$ regression equation
corresponds to measuring the correct distribution of $x_{k}$ conditional on $%
x_{R(k)}$. This is helpful expositionally and entails no loss of generality.

Given an objective distribution $p$ over $x_{1},...,x_{n}$ and a DAG $G$,
define the Bayesian-network factorization formula:%
\begin{equation*}
p_{G}(x_{1},...,x_{n})=\mathop{\displaystyle \prod }\limits_{k=1}^{n}p(x_{k}%
\mid x_{R(k)})
\end{equation*}%
We say that $p$ is consistent with $G$ if $p_{G}=p$.

By Koller and Friedman (2009, Ch. 7), when $p$ is multivariate normal, $%
p_{G} $ is reduced to the estimated joint distribution as described in
Section 2. In particular, we can use $p_{G}$ to calculate the estimated
marginal of $x_{k}$ for any $k$:%
\begin{equation*}
p_{G}(x_{k})=\mathop{\displaystyle \int}\limits_{(x_{j})_{j<k}}%
\mathop{\displaystyle \prod }\limits_{j\leq k}p(x_{j}\mid x_{R(j)})
\end{equation*}%
Likewise, the induced estimated distribution of $x_{n}$ conditional on $%
x_{1} $ is%
\begin{equation}
p_{G}(x_{n}\mid x_{1})=\mathop{\displaystyle \int}\limits_{x_{2},...,x_{n-1}}%
\mathop{\displaystyle \prod }\limits_{k\in K}p(x_{k}\mid x_{R(k)})
\end{equation}%
This conditional distribution, together with the marginals $p_{G}(x_{1})$
and $p_{G}(x_{n})$, induce the estimated correlation coefficient $\hat{\rho}%
_{1n}$ given by (\ref{estimated_correlation}).

Because we take $p$ to be multivariate normal, the constraint that $p_{G}$
does not distort the mean and variance of individual variables is equivalent
to the requirement that the estimated marginal distribution $(p_{G}(x_{k}))$
coincides with the objective marginal distribution of $x_{k}$. This
constraint necessarily holds if $G$ is perfect. Furthermore, when $G$ is
perfect, $p_{G}(x_{C})\equiv p(x_{C})$ for every clique $C$ in $G$ (see
Spiegler (2017)).

\subsection{The Proof}

Our first step is to establish that for generic $p$, perfection is necessary
for the correct-marginal constraint.\medskip

\begin{lemma}
\label{lemma1}Let $n\geq 3$ and suppose that $G$ is imperfect. Then, there
exists $k\in \{3,...,n\}$ such that $Var_{G}(x_{k})\neq 1$ for almost all
correlation submatrices $(\rho _{ij})_{i,j=1,...,k-1}$ (and therefore, for
almost all correlation matrices $(\rho _{ij})_{i,j=1,...,n}$).
\end{lemma}

\begin{proof}
Recall that we list the variables $x_{1},...,x_{n}$ such that $R(i)\subseteq
\{1,...,i-1\}$ for every $i$. Consider the lowest $k$ for which $R(k)$ is
not a clique. This means that there exist two nodes $h,l\in R(k)$ that are
unlinked in $G$, whereas for every $k^{\prime }<k$ and every $h^{\prime
},l^{\prime }\in R(k^{\prime })$, $h^{\prime }$ and $l^{\prime }$ are linked
in $G$.

Our goal is to show that $Var_{G}(x_{k})\neq 1$ for almost all correlation
submatrices $(\rho _{ij})_{i,j=1,...,k-1}$. Since none of the variables $%
x_{k+1},...,x_{n}$ appear in the equations for $x_{1},...,x_{k}$, we can
ignore them and treat $x_{k}$ as the terminal node in $G$ without loss of
generality, such that $G$ is defined over the nodes $1,...,k$, and $p$ is
defined over the variables $x_{1},...,x_{k}$.

Let $(\hat{\rho}_{ij})_{i,j=1,...,k-1}$ denote the correlation matrix over $%
x_{1},...,x_{k-1}$ induced by $p_{G}$ - i.e., $\hat{\rho}_{ij}$ is the
estimated correlation between $x_{i}$ and $x_{j}$, whereas $\rho _{ij}$
denotes their true correlation. By assumption, the estimated marginals of $%
x_{1},...,x_{k-1}$ are correct, hence $\hat{\rho}_{ii}=1$ for all $%
i=1,...,k-1$.

Furthermore, observe that in order to compute $\hat{\rho}_{ij}$ over $%
i,j=1,...,k-1$, we do not need to know the value of $\rho _{hl}$ (i.e. the
true correlation between $x_{h}$ and $x_{l}$). To see why, note that $(\hat{%
\rho}_{ij})_{i,j=1,...,k-1}$ is induced by $(p_{G}(x_{1},...,x_{k-1}))$.
Each of the terms in the factorization formula for $p_{G}(x_{1},...,x_{k-1})$
is of the form $p(x_{i}\mid x_{R(i)})$, $i=1,...,k-1$. To compute this
conditional probability, we only need to know $(\rho _{jj^{\prime
}})_{j,j^{\prime }\in \{i\}\cup R(i)}$. By the definition of $k$, $h$ and $l$%
, it is impossible for both $h$ and $l$ to be included in $\{i\}\cup R(i)$.
Therefore, we can compute $(\hat{\rho}_{ij})_{i,j=1,...,k-1}$ without
knowing the true value of $\rho _{hl}$. We will make use of this observation
toward the end of this proof.

The equation for $x_{k}$ is%
\begin{equation}
x_{k}=\sum_{i\in R(k)}\beta _{ik}x_{i}+\varepsilon _{k}  \label{reg_eq}
\end{equation}%
Let $\beta $ denote the vector $(\beta _{ik})_{i\in R(k)}$. Let $A$ denote
the correlation sub-matrix $(\rho _{ij})_{i,j\in R(k)}$ that fully
characterizes the objective joint distribution $(p(x_{R(k)}))$. Then, the
objective variance of $x_{k}$ can be written as%
\begin{equation}
Var(x_{k})=1=\beta ^{T}A\beta +\sigma ^{2}  \label{var_true}
\end{equation}%
where $\sigma ^{2}=Var(\varepsilon _{k})$.

In contrast, the estimated variance of $x_{k}$, denoted $Var_{G}(x_{k})$,
obeys the equation%
\begin{equation}
Var_{G}(x_{k})=\beta ^{T}C\beta +\sigma ^{2}  \label{var_estimated}
\end{equation}%
where $C$ denotes the correlation sub-matrix $(\hat{\rho}_{ij})_{i,j\in
R(k)} $ that characterizes $(p_{G}(x_{R(k)}))$. In other words, the
estimated variance of $x_{k}$ is produced by replacing the true joint
distributed of $x_{R(k)}$ in the regression equation for $x_{k}$ with its
estimated distribution (induced by $p_{G}$), without changing the values of $%
\beta $ and $\sigma ^{2}$.

The undistorted-marginals constraint requires $Var_{G}(x_{k})=1$. This
implies the equation%
\begin{equation}
\beta ^{T}A\beta =\beta ^{T}C\beta  \label{equation}
\end{equation}
We now wish to show that this equation fails for generic $(\rho
_{ij})_{i,j=1,...,k-1}$.

For any subsets $B,B^{\prime }\subset \{1,...,k-1\}$, use $\Sigma _{B\times
B^{\prime }}$ to denote the submatrix of $(\hat{\rho}_{ij})_{i,j=1,...,k-1}$
in which the selected set of rows is $B$ and the selected set of columns is $%
B^{\prime }$. By assumption, $h,l\in R(k)$ are unlinked. This means that
according to $G$, $x_{h}\perp x_{l}\mid x_{M}$, where $M\subset
\{1,...,k-1\}-\{h,l\}$. Therefore, by Drton et al. (2008, p. 67),%
\begin{equation}
\Sigma _{\{h\}\times \{l\}}=\Sigma _{\{h\}\times M}\Sigma _{M\times
M}^{-1}\Sigma _{M\times \{l\}}  \label{CI_equation}
\end{equation}%
Note that equation (\ref{CI_equation}) is precisely where we use the
assumption that $G$ is imperfect. If $G$ were perfect, then all nodes in $%
R(k)$ would be linked and therefore we would be unable to find a pair of
nodes $h,l\in R(k)$ that necessarily satisfies (\ref{CI_equation}).

The L.H.S of (\ref{CI_equation}) is simply $\hat{\rho}_{hl}$. The R.H.S of (%
\ref{CI_equation}) is induced by $p_{G}(x_{1},...,x_{k-1})$. As noted
earlier, this distribution is pinned down by $G$ and the entries in $(\rho
_{ij})_{i,j=1,...,k-1}$ except for $\rho _{hl}$. That is, if we are not
informed of $\rho _{hl}$ but we are informed of all the other entries in $%
(\rho _{ij})_{i,j=1,...,k-1}$, we are able to pin down the R.H.S of (\ref%
{CI_equation}).

Now, when we draw the objective correlation submatrix $(\rho
_{ij})_{i,j=1,...,k-1}$ at random, we can think of it as a two-stage
lottery. In the first stage, all the entries in this submatrix except $\rho
_{hl}$ are drawn. In the second stage, $\rho _{hl}$ is drawn. The only
constraint in each stage of the lottery is that $(\rho
_{ij})_{i,j=1,...,k-1} $ has to be positive-semi-definite and have $1$'s on
the diagonal. Fix the outcome of the first stage of this lottery. Then, it
pins down the R.H.S of (\ref{CI_equation}). In the lottery's second stage,
there is (for a generic outcome of the lottery's first stage) a continuum of
values that $\rho _{hl}$ could take for which $(\rho _{ij})_{i,j=1,...,k-1}$
will be positive-semi-definite. However, there is only value of $\rho _{hl}$
that will coincide with the value of $\hat{\rho}_{hl}$ that is given by the
equation (\ref{CI_equation}). We have thus established that $A\neq C$ for
generic $(\rho _{ij})_{i,j=1,...,k-1}$.

Recall once again that we can regards $\beta $ as a parameter of $p$ that is
independent of $A$ (and therefore of $C$ as well), because $A$ describes $%
(p(x_{R(k)}))$ whereas $\beta ,\sigma ^{2}$ characterize $(p(x_{k}\mid
x_{R(k)}))$. Then, since we can assume $A\neq C$, (\ref{equation}) is a
non-tautological quadratic equation of $\beta $ (because we can construct
examples of $p$ that violate it). By Caron and Traynor (2005), it has a
measure-zero set of solutions $\beta $. We conclude that the constraint $%
Var_{G}(x_{k})=1$ is violated by almost every $(\rho _{ij})$.\medskip
\end{proof}

\begin{corollary}
For almost every $(\rho _{ij})$, if a DAG $G$ satisfies $E_{G}(x_{k})=0$ and 
$Var_{G}(x_{k})=1$ for all $k=1,...,n$, then $G$ is perfect.
\end{corollary}

\begin{proof}
By Lemma \ref{lemma1}, for every imperfect DAG $G$, the set of covariance
matrices $(\rho _{ij})$ for which $p_{G}$ preserves the mean and variance of
all individual variables has measure zero. The set of imperfect DAGs over $%
\{1,...,n\}$ is finite, and the finite union of measure-zero sets has
measure zero as well. It follows that for almost all $(\rho _{ij})$, the
property that $p_{G}$ preserves the mean and variance of individual
variables is violated unless $G$ is perfect.\medskip
\end{proof}

The next step is based on the following definition.\medskip

\begin{definition}
A DAG $(N,R)$ is linear if $1$ is the unique ancestral node, $n$ is the
unique terminal node, and $R(i)$ is a singleton for every non-ancestral
node.\medskip
\end{definition}

A linear DAG is thus a causal chain $1\rightarrow \cdots \rightarrow n$.
Every linear DAG is perfect by definition.\medskip

\begin{lemma}
\label{lemma2}For every Gaussian distribution with correlation matrix $\rho $
and non-linear perfect DAG $G$ with $n$ nodes, there exists a Gaussian
distribution with correlation matrix $\rho ^{\prime }$ and a linear DAG $%
G^{\prime }$ with weakly fewer nodes than $G$, such that $\rho _{1n}=\rho
_{1n}^{\prime }$ and the false correlation induced by $G^{\prime }$ on $\rho
^{\prime }$ is exactly the same as the false correlation induced by $G$ on $%
\rho $: $cov_{G^{\prime }}(x_{1},x_{n})=cov_{G}(x_{1},x_{n})$.
\end{lemma}

\begin{proof}
The proof proceeds in two main steps.\medskip

\noindent \textit{Step 1: Deriving an explicit form for the false
correlation using an auxiliary \textquotedblleft cluster
recursion\textquotedblright\ formula}

\noindent The following is standard material in the Bayesian-network
literature. For any distribution $p_{G}(x)$ corresponding to a perfect DAG,
we can rewrite the distribution as if it factorizes according to a tree
graph, where the nodes in the tree are the maximal cliques of $G$. This tree
satisfies the \emph{running intersection property} (Koller and Friedman
(2009, p. 348)): If $i\in C,C^{\prime }$ for two tree nodes, then $i\in
C^{\prime \prime }$ for every $C^{\prime \prime }$ along the unique tree
path between $C^{\prime }$ and $C^{\prime \prime }$.\ Such a tree graph is
known as the \textquotedblleft \textit{junction tree}\textquotedblright\
corresponding to $G$ and we can write the following \textquotedblleft
cluster recursion\textquotedblright\ formula (Koller and Friedman (2009, p.
363)):%
\begin{equation*}
p_{G}(x)=p_{G}(x_{C_{r}})\prod_{i}p_{G}(x_{C_{i}}|x_{C_{r(i)}})=p(x_{C_{r}})%
\prod_{i}p(x_{C_{i}}|x_{C_{r(i)}})
\end{equation*}%
where $C_{r}$ is an arbitrary selected root clique node and $C_{r(i)}$ is
the upstream neighbor of clique $i$ (the one in the unique path from $C_{i}$
to the root $C_{r}$). The second equality is due to the fact that $G$ is
perfect, hence $p_{G}(x_{C})\equiv p(x_{C})$ for every clique $C$ of $G$.

Let $C_{1},C_{K}\in \mathcal{C}$ be two cliques that include the nodes $1$
and $n$, respectively. Furthermore, for a given junction tree representation
of the DAG, select these cliques to be minimally distant from each other -
i.e., $1,n\notin C$ for every $C$ along the junction-tree path between $%
C_{1} $ and $C_{K}$. We now derive an upper bound on $K$. Recall the running
intersection property: If $i\in C_{j},C_{k}$ for some $1\leq j<k\leq K$,
then $i\in C_{h}$ for every $h$ between $k$ and $j$. Since the cliques $%
C_{1},...,C_{K}$ are maximal, it follows that every $C_{k}$ along the
sequence must introduce at least one new element $i\notin \cup _{j<k}C_{j}$
(in particular, $C_{1}$ includes some $i>1$). As a result, it must be the
case that $K\leq n-1$. Furthermore, since $G$ is assumed to be non-linear,
the inequality is $strict$, because at least one $C_{k}$ along the sequence
must contain at least three elements and therefore introduce at least $two$
new elements. Thus, $K\leq n-2$.

Since $p_{G}$ factorizes according to the junction tree, it follows that the
distribution over the variables covered by the cliques along the path from $%
C_{1}$ to $C_{K}$ factorize according to a linear DAG $1\rightarrow
C_{1}\rightarrow \cdots \rightarrow C_{K}\rightarrow n$, as follows:%
\begin{equation}
p_{G}(x_{1},x_{C_{1}},...,x_{C_{K}},x_{n})=p(x_{1})%
\prod_{k=1}^{K}p(x_{C_{k}}|x_{C_{k-1}})p(x_{n}|x_{C_{K}})
\label{junctiontree}
\end{equation}%
where $C_{0}=\{1\}$. The length of this linear DAG is $K+2\leq n$. While
this factorization formula superficially completes the proof, note that the
variables $x_{C_{k}}$ are typically $multivariate$ normal variables, whereas
our objective is to show that we can replace them with scalar (i.e.
univariate) normal variables without changing $cov_{G}(x_{1},x_{n})$.

Recall that we can regard $p$ as a multivariate normal distribution without
loss of generality. Furthermore, under such a distribution and any two
subsets of variables $C,C^{\prime }$, the distribution of $x_{C}$
conditional on $x_{C^{\prime }}$ can be written $x_{C}=Ax_{C^{\prime }}+\eta 
$, where $A$ is a matrix that depends on the means and covariances of $p$,
and $\eta $ is a zero-mean vector that is uncorrelated with $x_{C^{\prime }}$%
. Applying this property to the junction tree, we can describe $%
p_{G}(x_{1},x_{C_{1}},...,x_{C_{K}},x_{n})$ via the following recursion:%
\begin{eqnarray}
x_{1} &\sim &N(0,1)  \label{recursion} \\
x_{C_{1}} &=&A_{1}x_{1}+\eta _{1}  \notag \\
&&\vdots  \notag \\
x_{C_{k}} &=&A_{k}x_{C_{k-1}}+\eta _{k}  \notag \\
&&\vdots  \notag \\
x_{C_{K}} &=&A_{K}x_{C_{K-1}}+\eta _{K}  \notag \\
x_{n} &=&A_{K+1}x_{C_{K}}+\eta _{n}  \notag
\end{eqnarray}%
where each equation describes an objective conditional distribution - in
particular, the equation for $x_{C_{k}}$ describes $%
(p(x_{C_{k}}|x_{C_{k-1}}))$. The matrices $A_{k}$ are functions of the
vectors $\beta _{i}$ in the original recursive model. The $\eta _{k}$'s are
all zero mean and uncorrelated with the explanatory variables $x_{C_{k-1}}$,
such that $E(x_{C_{k}}|x_{C_{k-1}})=A_{k}x_{C_{k-1}}$. Furthermore,
according to $p_{G}$ (i.e. the analyst's estimated model), each $x_{k}$
(with $k>1$) is conditionally independent of $x_{1},...,x_{k-1}$ given $%
x_{R(k)}$. Since the junction-tree factorization (\ref{junctiontree})
represents exactly the same distribution $p_{G}$, this means that every $%
\eta _{k}$ is uncorrelated with all other $\eta _{j}$'s as well as with $%
x_{1},...,x_{C_{k-2}}$. Therefore,%
\begin{equation*}
E_{G}(x_{1}x_{n})=A_{K+1}A_{K}\cdots A_{1}
\end{equation*}%
Since $p_{G}$ preserves the marginals of individual variables, $%
Var_{G}(x_{k})=1$ for all $k$. In particular $%
Var_{G}(x_{1})=Var_{G}(x_{n})=1 $ Then,%
\begin{equation*}
\rho _{G}(x_{1}x_{n})=A_{K+1}A_{K}\cdots A_{1}
\end{equation*}%
\medskip

\noindent \textit{Step 2: Defining a new distribution over scalar variables}

\noindent For every $k$, define the variable%
\begin{equation*}
z_{k}=\left( A_{K+1}A_{K}\cdots A_{k+1}\right) x_{C_{k}}=\alpha _{k}x_{C_{k}}
\end{equation*}

Plugging the recursion (\ref{recursion}), we obtain a recursion for $z$:%
\begin{eqnarray*}
z_{k} &=&\alpha _{k}x_{C_{k}} \\
&=&\alpha _{k}(A_{k}x_{C_{k-1}}+\eta _{k}) \\
&=&z_{k-1}+\alpha _{k}\eta _{k}
\end{eqnarray*}%
Given that $p$ is taken to be multivariate normal, the equation for $z_{k}$
measures the objective conditional distribution $(p_{G}(z_{k}\mid z_{k-1}))$%
. Since $p_{G}$ does not distort the objective distribution over cliques, $%
(p_{G}(z_{k}\mid z_{k-1}))$ coincides with $(p(z_{k}\mid z_{k-1}))$. This
means that an analyst who fits a recursive model given by the linear DAG $%
G^{\prime }:x_{1}\rightarrow z_{1}\rightarrow \cdots \rightarrow
z_{K}\rightarrow x_{n}$ will obtain the following estimated model, where
every $\varepsilon _{k}$ is a zero-mean scalar variable that is assumed by
the analyst to be uncorrelated with the other $\varepsilon _{j}$'s as well
as with $z_{1},...,z_{k}$ (and as before, the assumption holds automatically
for $z_{k}$ but is typically erroneous for $z_{j}$, $j<k$):%
\begin{eqnarray*}
x_{1} &\sim &N(0,1) \\
z_{1} &=&\alpha _{1}A_{1}x_{1}+\varepsilon _{2} \\
&&\vdots \\
z_{k+1} &=&z_{k}+\varepsilon _{k+1} \\
&&\vdots \\
x_{n} &=&z_{K}+\varepsilon _{n}
\end{eqnarray*}%
Therefore, $E_{G^{\prime }}(x_{1},x_{n})$ is given by%
\begin{equation*}
E_{G^{\prime }}(x_{1}x_{n})=A_{K+1}A_{K}\cdots A_{1}
\end{equation*}%
Since $G^{\prime }$ is perfect, $Var_{G^{\prime }}(x_{n})=1$, hence%
\begin{equation*}
\rho _{G^{\prime }}(x_{1}x_{n})=A_{K+1}A_{K}\cdots A_{1}=\rho
_{G}(x_{1}x_{n})
\end{equation*}

We have thus reduced our problem to finding the largest $\hat{\rho}_{1n}$
that can be attained by a linear DAG $G:1\rightarrow \cdots \rightarrow n$
of length $n$ at most.\medskip
\end{proof}

To solve the reduced problem we have arrived at, we first note that%
\begin{equation}
\hat{\rho}_{1n}=\prod_{k=1}^{n-1}\rho _{k,k+1}  \label{chain-rule-eq}
\end{equation}%
Thus, the problem of maximizing $\hat{\rho}_{1n}$ is equivalent to
maximizing the product of terms in a symmetric $n\times n$ matrix, subject
to the constraint that the matrix is positive semi-definite, all diagonal
elements are equal to one, and the $(1,n)$ entry is equal to $r$:%
\begin{equation*}
\rho _{1n}^{\ast }=\max_{\substack{ \rho _{ij}=\rho _{ji}\text{ for all }i,j 
\\ (\rho _{ij})\text{ is P.S.D}  \\ \rho _{ii}=1\text{ for all }i  \\ \rho
_{1n}=r}}\text{ \ }\prod_{i=1}^{n-1}\rho _{i,i+1}
\end{equation*}%
Note that the positive semi-definiteness constraint is what makes the
problem nontrivial. We can arbitrarily increase the value of the objective
function by raising off-diagonal terms of the matrix, but at some point this
will violate positive semi-definiteness. Since positive semi-definiteness
can be rephrased as the requirement that $(\rho _{ij})=AA^{T}$ for some
matrix $A$, we can rewrite the constrained maximization problem as follows:%
\begin{equation}
\rho _{1n}^{\ast }=\max_{\substack{ a_{i}^{T}a_{i}=1\text{ for all }i  \\ %
a_{1}^{T}a_{n}=r}}\prod_{i=1}^{n-1}a_{i}a_{i+1}^{T}
\label{shortest-path-sphere-eq}
\end{equation}

Denote $\alpha =\arccos r$. Since the solution to (\ref%
{shortest-path-sphere-eq}) is invariant to a rotation of all vectors $a_{i}$%
, we can set%
\begin{eqnarray*}
a_{1} &=&e_{1} \\
a_{n} &=&e_{1}\cos \alpha +e_{2}\sin \alpha
\end{eqnarray*}%
without loss of generality. Note that $a_{1},a_{n}$ are both unit norm and
have dot product $r$. Thus, we have eliminated the constraint $%
a_{1}^{T}a_{n}=r$ and reduced the variables in the maximization problem to $%
a_{2},...,a_{n-1}$.

Now consider some $k=2,...,n-1$. Fix $a_{j}$ for all $j\neq k$, and choose $%
a_{k}$ to maximize the objective function. As a first step, we show that $%
a_{k}$ must be a linear combination of $a_{k-1},a_{k+1}$. To show this, we
write $a_{k}=u+v$, where $u,v$ are orthogonal vectors, $u$ is in the
subspace spanned by $a_{k-1},a_{k+1}$ and $v$ is orthogonal to the subspace.
Recall that $a_{k}$ is a unit-norm vector, which implies that 
\begin{equation}
\Vert u\Vert ^{2}+\Vert v\Vert ^{2}=1  \label{norm-eq-1-eq}
\end{equation}

The terms in the objective function (\ref{shortest-path-sphere-eq}) that
depend on $a_{k}$ are simply $(a_{k-1}^{T}u)(a_{k+1}^{T}u)$. All the other
terms in the product do not depend on $a_{k}$, whereas the dot product
between $a_{k}$ and $a_{k=1},a_{k+1}$ is invariant to $v$: $%
a_{k-1}^{T}(u+v)=a_{k=1}^{T}u$.

Suppose that $v$ is nonzero. Then, we can replace $a_{k}$ with another
unit-norm vector $u/\Vert u\Vert $, such that $(a_{k-1}^{T}u)(a_{k+1}^{T}u)$
will be replaced by%
\begin{equation*}
\frac{(a_{k-1}^{T}u)(a_{k+1}^{T}u)}{\Vert u\Vert ^{2}}
\end{equation*}%
By (\ref{norm-eq-1-eq}) and the assumption that $v$ is nonzero, $\Vert
u\Vert <1$, hence the replacement is an improvement. It follows that $a_{k}$
can be part of an optimal solution only if it lies in the subspace spanned
by $a_{k-1},a_{k+1}$. Geometrically, this means that $a_{k}$ lies in the
plane defined by the origin and $a_{k-1},a_{k+1}$.

Having established that $a_{k},a_{k-1},a_{k+1}$ are coplanar, let $\alpha $
be the angle between $a_{k}$ and $a_{k-1}$, let $\beta $ be the angle
between $a_{k}$ and $a_{k+1}$, and let $\gamma $ be the (fixed) angle
between $a_{k-1}$ and $a_{k+1}$. Due to the coplanarity constraint, $\alpha
+\beta =\gamma $. Fixing $a_{j}$ for all $j\neq k$ and applying a
logarithmic transformation to the objective function, the optimal $a_{k}$
must satisfy%
\begin{equation*}
\log \cos (\alpha )+\log \cos (\gamma -\alpha )
\end{equation*}%
Differentiating this expression with respect to $\alpha $ and setting the
derivative to zero, we obtain $\alpha =\beta =\gamma /2$. Since this must
hold for \emph{any} $k=2,...,n-1$, we conclude that at the optimum, any $%
a_{k}$ lies on the plane defined by the origin and $a_{k-1},a_{k+1}$ and is
at the same angular distance from $a_{k-1},a_{k+1}$. That is, an optimum
must be a set of equiangular unit vectors on a great circle, equally spaced
between $a_{1}$ and $a_{n}$. The explicit formulas for these vectors are
given by (\ref{solution_xk}).


\begin{figure}[tbp]
\centerline{\includegraphics[width=0.5%
\columnwidth]{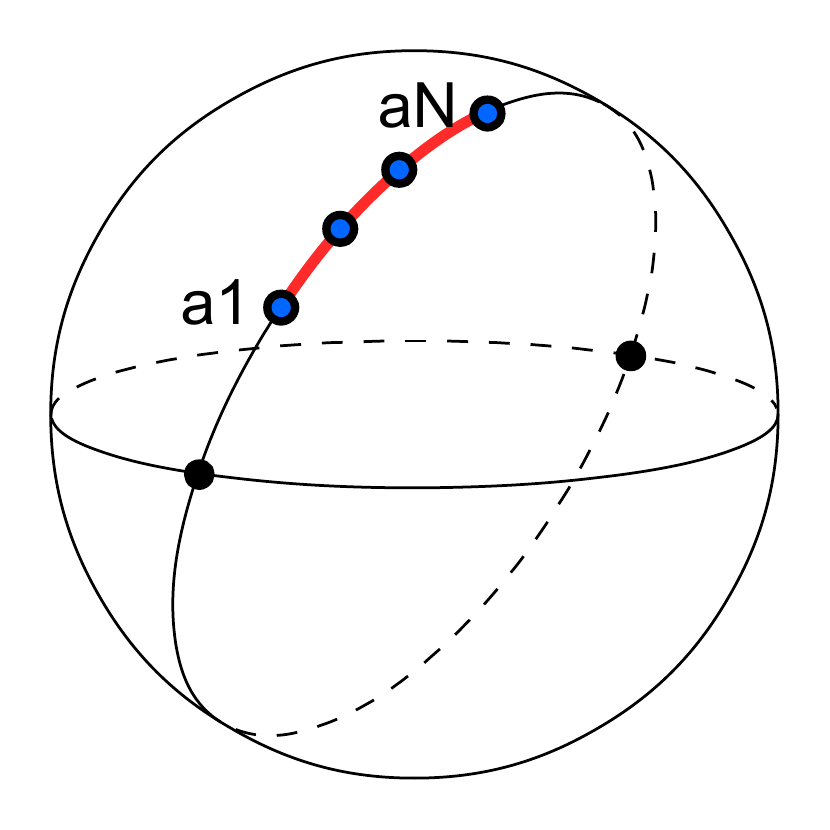}}
\caption{Geometric intuition for the proof.}
\end{figure}

The formula for the upper bound has a simple geometric interpretation
(illustrated by Figure 2). We are given two points on the unit $n$%
-dimensional sphere (representing $a_{1}$ and $a_{n}$) whose dot product is $%
r$, and we seek\ $n-2$ additional points on the sphere such that the
harmonic average of the successive points' dot product is maximal. Since the
dot product for points on the unit sphere decreases with the spherical
distance between them, the problem is akin to minimizing the average
distance between adjacent points. The solution is to place all the
additional points equidistantly on the great circle that connects $a_{1}$
and $a_{n}$.

Since by construction, every neighboring points $a_{k}$ and $a_{k+1}$ have a
dot product of $\cos \theta _{r,n}$, we have $\rho _{k,k+1}=\cos \theta
_{r,n}$, such that $\hat{\rho}_{1n}=(\cos \theta _{r,n})^{n-1}$. This
completes the proof.

\section{Conclusion}

This paper performed a worst-case analysis of misspecified recursive models.
We showed that within this class, model selection is a very powerful tool in
the hands of an opportunistic analyst: If we allow him to freely select a
moderate number of variables from a large pool, he can produce a very large
estimated correlation between two variables of interest. Furthermore, the
structure of his model allows him to interpret this correlation as a causal
effect. This is true even if the two variables are objectively independent,
or if their correlation is in the opposite direction. Imposing a bound on
the model's complexity (measured by its number of auxiliary variables) is an
important constraint on the analyst. However, the value of this bound decays
quickly, as even with one or two auxiliary variables the analyst can greatly
distort objective correlations.

Within our framework, several questions are left open. First, we do not know
whether Theorem \ref{thm1} would continue to hold if we replaced the
quantifier \textquotedblleft for almost every $p$\textquotedblright\ with
\textquotedblleft for every $p$\textquotedblright . Second, we do not know
how much bite the undistorted-variance constraint has in models with more
than one non-trivial equation. Third, we lack complete characterizations for
recursive models outside the linear-regression family (see our partial
characterization for models that involve binary variables in Appendix II).
Finally, it would be interesting to devise a sparse collection of
misspecification or robustness tests that would restrain our opportunistic
analyst.

Taking a broader perspective into the last question, our exercise suggests a
novel approach to the study of biased estimates due to misspecified models
in Statistics and Econometric Theory (foreshadowed by Spiess (2018)). Under
this approach, the analyst who employs a structural model for statistical or
causal analysis is viewed as a player in a game with his audience.
Researcher bias implies a conflict of interests between the two parties.
This bias means that the analyst's model selection is opportunistic. The
question is which strategies the audience can play (in terms of robustness
or misspecification tests it can demand) in order to mitigate errors due to
researcher bias, without rejecting too many valuable models.\bigskip \bigskip

\noindent {\LARGE Appendix: Uniform Binary Variables}\medskip

\noindent Suppose now that the variables $x_{1},...,x_{n}$ all take values
in $\{-1,1\}$, and restrict attention to the class of objective
distributions $p$ whose marginal on each variable is uniform - i.e., $%
p(x_{i}=1)=\frac{1}{2}$ for every $i=1,...,n$. As in our main model, fix the
correlation between $x_{1}$ and $x_{n}$ to be $r$ - that is,%
\begin{equation*}
\rho _{1n}=p(x_{n}=1\mid x_{1}=1)-p(x_{n}=1\mid x_{1}=-1)=r
\end{equation*}

The question of finding the distribution $p$ (in the above restricted
domain) and the DAG $G$ that maximize the induced $\hat{\rho}_{in}$ subject
to $p_{G}(x_{n})=\frac{1}{2}$ is generally open. However, when we fix $G$ to
be the linear DAG%
\begin{equation*}
1\rightarrow 2\rightarrow \cdots \rightarrow n
\end{equation*}%
we are able to find the maximal $\hat{\rho}_{1n}$. It makes sense to
consider this specific DAG, because it proved to be the one most conducive
to generating false correlations in the case of linear-regression models.

Given the DAG $G$ and the objective distribution $p$, the correlation
between $x_{i}$ and $x_{j}$ that is induced by $p_{G}$ is%
\begin{equation*}
\hat{\rho}_{ij}=p_{G}(x_{j}=1\mid x_{i}=1)-p_{G}(x_{j}=1\mid x_{i}=-1)
\end{equation*}%
Let $j>i$. Given the structure of the linear DAG, we can write%
\begin{equation}
p_{G}(x_{j}\mid x_{i})=\sum_{x_{i+1},...x_{j-1}}p(x_{i+1}\mid
x_{i})p(x_{i+2}\mid x_{i+1})\cdots p(x_{j}\mid x_{j-1})  \label{p_R}
\end{equation}

In particular, 
\begin{eqnarray}
p_{G}(x_{n} &\mid &x_{1})=\sum_{x_{2},...x_{n-1}}p(x_{2}\mid
x_{1})p(x_{3}\mid x_{2})\cdots p(x_{n}\mid x_{n-1})  \label{recursive} \\
&=&\sum_{x_{2}}p(x_{2}\mid x_{1})p_{G}(x_{n}\mid x_{2})  \notag
\end{eqnarray}%
Note that $p_{G}(x_{n}\mid x_{2})$ has the same expression that we would
have if we dealt with a linear DAG of length $n-1$, in which $2$ is the
ancestral node: $2\rightarrow \cdots \rightarrow n$. This observation will
enable us to apply an inductive proof to our result.\medskip

\begin{lemma}
For every $p$,%
\begin{equation*}
\hat{\rho}_{1n}=\rho _{12}\cdot \hat{\rho}_{2n}
\end{equation*}
\end{lemma}

\begin{proof}
Applying simple algebraic manipulation of (\ref{recursive}), $\hat{\rho}%
_{1n} $ is equal to%
\begin{eqnarray*}
&&\left[ p(x_{2}=1\mid x_{1}=1)-p(x_{2}=1\mid x_{1}=-1)\right] \left[
p_{G}(x_{n}=1\mid x_{2}=1)-p_{G}(x_{n}=1\mid x_{2}=-1)\right] \\
&=&\rho _{12}\cdot \hat{\rho}_{2n}\medskip
\end{eqnarray*}
\end{proof}

We can now derive an upper bound on $\hat{\rho}_{in}$ for the environment of
this appendix - i.e., the estimated model is a linear DAG, and the objective
distribution has uniform marginals over binary variables.$\medskip $

\begin{proposition}
For every $n$,%
\begin{equation*}
\hat{\rho}_{1n}\leq \left( 1-\frac{1-r}{n-1}\right) ^{n-1}
\end{equation*}
\end{proposition}

\begin{proof}
The proof is by induction on $n$. Let $n=2$. Then, $p_{G}(x_{2}\mid
x_{1})=p(x_{2}\mid x_{1})$, and therefore $\hat{\rho}_{12}=r$, which
confirms the formula.

Suppose that the claim holds for some $n=k\geq 2$. Now let $n=k+1$. Consider
the distribution of $x_{2}$ conditional on $x_{1},x_{n}$. Denote $\alpha
_{x_{1},x_{n}}=p(x_{2}=1\mid x_{1},x_{n})$. We wish to derive a relation
between $\rho _{12}$ and $\rho _{2n}$. Denote 
\begin{equation*}
q=\frac{1+r}{2}=p(x_{n}=1\mid x_{1}=1)=p(x_{n}=-1\mid x_{1}=-1)
\end{equation*}%
Then,%
\begin{eqnarray*}
p(x_{2} &=&1\mid x_{1}=1)=p(x_{n}=1\mid x_{1}=1)\cdot \alpha
_{1,1}+p(x_{n}=-1\mid x_{1}=1)\cdot \alpha _{1,-1} \\
&=&q\alpha _{1,1}+(1-q)\alpha _{1,-1}
\end{eqnarray*}%
Likewise,%
\begin{eqnarray*}
p(x_{2} &=&1\mid x_{1}=0)=p(x_{n}=1\mid x_{1}=0)\cdot \alpha
_{-1.1}+p(x_{n}=0\mid x_{1}=0)\cdot \alpha _{-1,-1} \\
&=&q\alpha _{-1,-1}+(1-q)\alpha _{-1,1}
\end{eqnarray*}%
The objective correlation between $x_{1}$ and $x_{2}$ is thus%
\begin{equation}
\rho _{12}=q(\alpha _{1,1}-\alpha _{-1,-1})+(1-q)(\alpha _{1,-1}-\alpha
_{-1,1})  \label{corr(x1,x2)}
\end{equation}%
Let us now turn to the joint distribution of $x_{n}$ and $x_{2}$. Because
the marginals on both $x_{2}$ and $x_{n}$ are uniform, $p(x_{n}\mid
x_{2})=p(x_{2}\mid x_{n})$. Therefore, we can obtain $\rho _{2n}$ in the
same manner that we obtained $\rho _{12}$:%
\begin{equation}
\rho _{2n}=q(\alpha _{1,1}-\alpha _{-1,-1})+(1-q)(\alpha _{-1,1}-\alpha
_{1,-1})  \label{corr(x2,xn)}
\end{equation}%
We have thus established a relation between $\rho _{12}$ and $\rho _{2n}$.

Recall that $\hat{\rho}_{2n}$ is the expression we would have for the linear
DAG $2\rightarrow \cdots \rightarrow n$ when $p(x_{2}=x_{n})=\tilde{q}$.
Therefore, by the inductive step,%
\begin{eqnarray}
\hat{\rho}_{1n} &=&\rho _{12}\cdot \hat{\rho}_{2n}  \label{induction_ineq} \\
&\leq &[q(\alpha _{1,1}-\alpha _{-1,-1})+(1-q)(\alpha _{1,-1}-\alpha
_{-1,1})]\cdot \left( 1-\frac{1-\rho _{2n}}{k-1}\right) ^{k-1}  \notag
\end{eqnarray}%
Both $\rho _{12}$ and $\rho _{2n}$ increase in $\alpha _{1,1}$ and decrease
in $\alpha _{-1,-1}$, such that we can set $\alpha _{1,1}=1$ and $\alpha
_{-1,-1}=0$ without lowering the R.H.S of (\ref{induction_ineq}). This
enables us to write 
\begin{equation*}
\rho _{12}=q+(1-q)(\alpha _{1,-1}-\alpha _{-1,1})
\end{equation*}%
such that%
\begin{equation*}
\rho _{2n}=1+r-\rho _{12}
\end{equation*}%
Therefore, we can transform (\ref{induction_ineq}) into%
\begin{equation*}
\hat{\rho}_{1n}\leq \max_{\rho _{12}}\quad \rho _{12}\cdot \left( 1-\frac{%
\rho _{12}-r}{k-1}\right) ^{k-1}
\end{equation*}%
The R.H.S is a straightforward maximization problem. Performing a
logarithmic transformation and writing down the first-order condition, we
obtain%
\begin{equation*}
\rho _{12}^{\ast }=1-\frac{1-r}{k}
\end{equation*}%
and%
\begin{equation*}
\left( 1-\frac{\rho _{12}^{\ast }-r}{k-1}\right) ^{k-1}=\left( 1-\frac{1-r}{k%
}\right) ^{k-1}
\end{equation*}%
such that%
\begin{equation*}
\hat{\rho}_{1n}\leq \left( 1-\frac{1-r}{k}\right) ^{k}
\end{equation*}%
which completes the proof.\medskip
\end{proof}

How does this upper bound compare with the Gaussian case? For illustration,
let $r=0$. Then, it is easy to see that for $n=3$, we obtain $\hat{\rho}%
_{13}=\frac{1}{3}$, which is below the value of $\frac{1}{2}$ we were able
to obtain in the Gaussian case. And as $n\rightarrow \infty $, $\hat{\rho}%
_{1n}\rightarrow 1/e$. That is, unlike the Gaussian case, the maximal false
correlation that the linear DAG can generate is bounded far away from one.

The upper bound obtained in this result is tight. The following is one way
to implement it. For the case $r=0$, take the exact same Gaussian
distribution over $x_{1},...,x_{n}$ that we used to implement the upper
bound in Theorem \ref{thm1}, and now define the variable $y_{k}=sign(x_{k})$
for each $k=1,...,n$. Clearly, each $y_{k}\in \{-1,1\}$ and $%
p(y_{k}=1)=p(y_{k}=-1)=\frac{1}{2}$ since each $x_{k}$ has zero mean. To
find the correlations between different $y_{k}$ variables, we use the
following lemma.\medskip

\begin{lemma}
Let $w_{1},w_{2}$ be two unit vectors in $R^{2}$ and let $z$ be a
multivariate Gaussian with zero mean and unit covariance. Then,%
\begin{equation*}
E(sign(w_{1}^{T}z)sign(w_{2}^{T}z))=1-\frac{2\theta }{\pi }
\end{equation*}%
where $\theta $ is the angle between the two vectors.
\end{lemma}

\begin{proof}
This follows from the fact that the product $%
sign(w_{1}^{T}z)sign(w_{2}^{T}z) $\ is equal to $1$ whenever $z$ is on the
same side of the two hyperplanes defined by $w_{1}$ and $w_{2}$, and $-1$
otherwise. Since the Gaussian distribution of $z$ is circularly symmetric,
the probability that $z$ lies on the same side of the two hyperplanes
depends only on the angle between them.\medskip
\end{proof}

Returning to the definition of the Gaussian distribution over $%
x_{1},...,x_{n}$ that we used to implement the upper bound in Theorem \ref%
{thm1}, we see that in the case of $r=0$, the angle between $w_{1}$ and $%
w_{n}$ will be $\frac{\pi }{2}$, so that by the above lemma, $y_{1}$ and $%
y_{n}$ will be uncorrelated. At the same time, the angle between any $w_{k}$
and $w_{k-1}$ is by construction $\frac{\pi }{2}\frac{1}{n-1}$ because the
vectors were chosen at equal angles along the great circle. Substituting
this angle into the lemma, we obtain that the correlation between $y_{k}$
and $y_{k-1}$ is $1-\frac{1}{n-1}$.

For the case where $r\neq 0$, the same argument holds, except that we need
to choose the original vectors $w_{1},w_{n}$ so that the correlation between 
$y_{1}$ and $y_{n}$ will be $r$ (these will not be the same vectors that
give a correlation of $r$ between the Gaussian variables $x_{1}$ and $x_{n}$%
) and then choose the rest of the vectors at equal angles along the great
circle. By applying the lemma again, we obtain that the angle between $y_{k}$
and $y_{k-1}$ is $1-\frac{1-r}{n-1}$, which again attains the upper bound.

This method of implementing the upper bound also explains why false
correlations are harder to generate in the uniform binary case, compared
with the case of linear-regression models. The variable $y_{k}$ is a
coarsening of the original Gaussian variable $x_{k}$. It is well-known that
when we coarsen Gaussian variables, we weaken their mutual correlation.
Therefore, the correlation between any consecutive variables $y_{k},y_{k+1}$
in the construction for the uniform binary case is lower than the
corresponding correlation in the Gaussian case. As a result, the maximal
correlation that the model generates is also lower.

The obvious open question is whether the restriction to linear DAGs entails
in a loss of generality. We conjecture that in the case of uniform binary
variables, a non-linear perfect DAG can generate larger false correlations
for sufficiently large $n$.

\end{document}